# Stability of Explainable Recommendation

Sairamvinay Vijayaraghavan
saivijay@ucdavis.edu
University of California Davis
Davis, CA, USA

Prasant Mohapatra
pmohapatra@ucdavis.edu
University of California, Davis
Davis, USA

## ABSTRACT

Explainable Recommendation has been gaining attention over the last few years in industry and academia. Explanations provided along with recommendations in a recommender system framework have many uses: particularly reasoning why a suggestion is provided and how well an item aligns with a user's personalized preferences. Hence, explanations can play a huge role in influencing users to purchase products. However, the reliability of the explanations under varying scenarios has not been strictly verified from an empirical perspective. Unreliable explanations can bear strong consequences such as attackers leveraging explanations for manipulating and tempting users to purchase target items that the attackers would want to promote. In this paper, we study the vulnerability of existent feature-oriented explainable recommenders, particularly analyzing their performance under different levels of external noises added into model parameters. We conducted experiments by analyzing three important state-of-the-art (SOTA) explainable recommenders when trained on two widely used e-commerce based recommendation datasets of different scales. We observe that all the explainable models are vulnerable to increased noise levels. Experimental results verify our hypothesis that the ability to explain recommendations does decrease along with increasing noise levels and particularly adversarial noise does contribute to a much stronger decrease. Our study presents an empirical verification on the topic of robust explanations in recommender systems which can be extended to different types of explainable recommenders in RS.

## 1 INTRODUCTION

Explainability of Recommender Systems (RS) is an important field which studies methods that learn why a recommendation is suggested by a model for a user [14, 15, 39]. The explanations provided improve the transparency of the system, by revealing more about the predicted outcome as in how does the model learn personalized preferences for every user [37, 45]. Moreover, explanations provided within a RS framework can directly appeal to a user and even influence them to purchase an item if it is very well explained (by providing very detailed information associated with the recommendation) as to why it is recommended to the user [26]. Additionally, explanations can be leveraged for detecting anomalies in certain systems [6], such as in graph neural networks [13]. Thus, explanations provided along with recommendations must be reliable and unchanging under varying scenarios, however the current existing explainable recommenders are typically vulnerable towards external attacks and hence provide unstable explanations.

While there has been a lot of work done in improving the explainability of the model, there has not been much attention drawn towards studying the robustness of explanations provided by a recommender during varying circumstances [14, 15]. Explainable systems that are prone to attacks can provide an easy outlet for attackers with malicious intent to achieve their objectives. For example in figure 1, we present this consequence within an example of cellphone recommendation in e-commerce websites, where the attackers (say a particular mobile brand's manufacturer) can promote a target mobile phone (belonging to that particular brand of the attacker, in this case, item $C$ in red) by deliberately manipulating the associated explanations for a user's personalized recommendations. This form of manipulation can be done by adjusting the feature scores- battery and/or screen quality of the item to align more with a current user's (say $U$) interests which is utilized as explanations, to match the user's preferences. This could attract the users to interact more with the target item and hence grab their interest away from the much more relevant items (items $A$ and $B$ in green). Hence, this could provide a different representation from the original characteristics of the target item, thus tricking consumers to purchase them and hence achieving the attacker's objective.

In this work, we present an empirical research study on the existent explainable recommender models by exploring the global explainability of the model under noisy environments. For a simplistic empirical verification, we chose feature-aware aspect-based explainable recommenders and for attacks, we chose a white-box-based attack that involves perturbing the model parameters. Our main contribution of this work is to empirically confirm that explainable methods in RS are indeed vulnerable to external security-based attacks which could possibly deteriorate the quality of the explanations. Our contributions are: (1) We focus on the intrinsic learning of the model and the subsequent explanations, thus emphasizing on the need for reliable recommendations and stable explanations together; and (2) We empirically verify the importance of reliable explanations from the standpoint of consumers as well as developers in RS, therein identifying an important field of work in trustworthy AI.

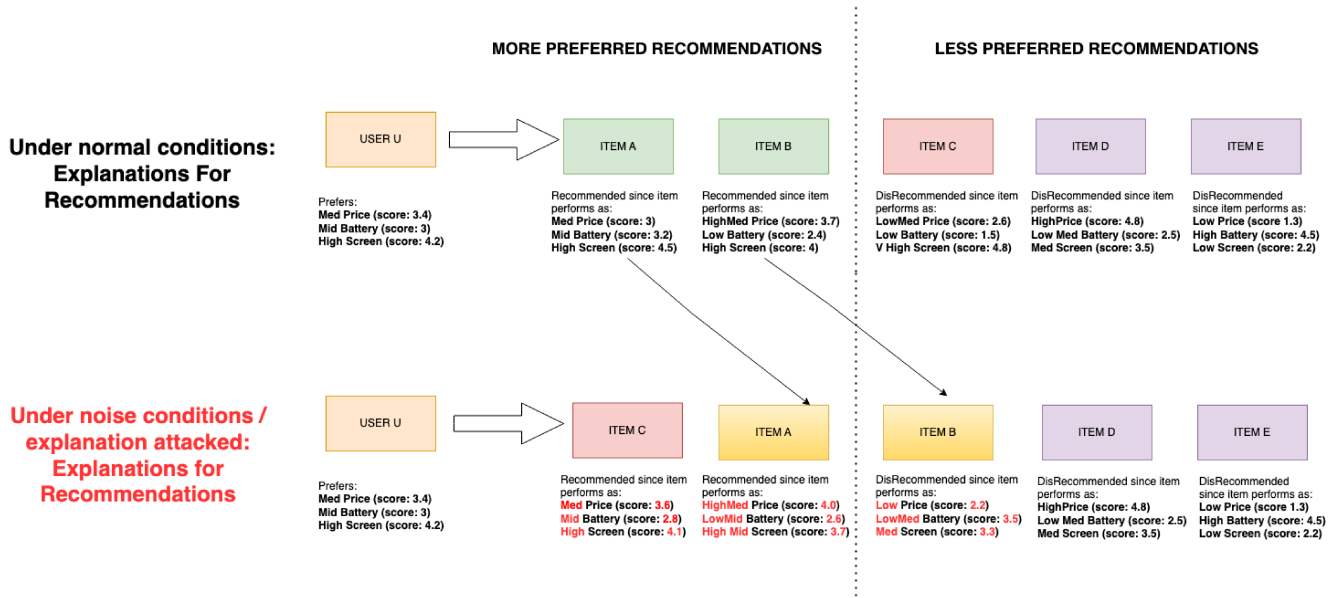

Figure 1: Motivation for reliable explanations

## 2 RELATED WORK
## 2.1 Explainable Recommendation

Explainability in RS is a broad research topic that has been covered by providing explanations in many forms. Explanations are provided typically in the form of generated textual explanations [24, 40, 41], or feature-level explanations [9, 29, 38]. Recently knowledge graphs have been extensively utilized in recommender systems to provide much more insight into how much did peers' influence plays a role in recommendations [10, 21, 30]. There is a much more comprehensive survey of all the explainable recommendation strategies found in [45]. We attain our focus more on feature-based explanations from user-provided text reviews since they are the easiest form of explanations for developers as well as consumers to understand and interpret. Zhang et al. [47] perform sentiment analysis on text reviews from e-commerce websites in extracting the important salient aspects along with the adjective and the sentiment in which they are expressed while purchasing an item. These components are utilized in different ways for user-preference modeling and explaining the personalized recommendations for a user. In [46], Zhang et al. pioneer a novel method of decomposing user-feature and item-feature matrices in such a way that the sentiment polarity of features is captured. The same structure of feature engineering was leveraged in these works also [8, 16, 23, 35, 38, 48]. While all the above methods have an aim of improving the explainability and the recommendation accuracy of the system, our work differs from the perspective that we analyze explainability under different noise scenarios and analyze the strength of the explanations generated.

## 2.2 Stability of Explanations

In recent years, machine learning has progressed in the research of explainable AI, by considering the different aspects of explainability such as faithfulness, fairness, and particularly stability [1–3, 5]. In [22], Kim et al. propose that the loss gradients which represent the model interpretability improves as adversarial training is included within deep neural networks (DNN). Even in [12], Etmann et al. verify that models which are more robust against adversarial attacks lead to much more interpretable saliency maps. Recent works confirm the origin of this phenomenon in existing gradient-based interpretation methods [28, 32]. For example, Ghorbani et al. [17] learn perturbations using gradient sign attacks added into the inputs such that they have different explanations provided by a gradient-based interpretation method called DeepLIFT [31]) despite the explained samples belonging to the same prediction class. They imply that there is a geometric connotation to the above phenomenon. Similarly, Dombrowski et al. [11] confirm that the decision functions of neural networks possess a large curvature (verified via Hessian analysis) which imply that similar-appearing images could possibly be interpreted differently when small perturbations are introduced into the images. The impetus for perturbing explanations has typically been from the standpoint of inputs, with a strong claim that similar inputs must possess similar interpretations [1, 4, 33, 44]. Our study also intends to verify the robustness of explainable models but strictly within the realm of RS.

## 3 PROBLEM STATEMENT
### 3.1 Feature-aware Recommender System

Let $U$ be the set of users and $V$ be the set of items of a dataset $D$. Following the sentiment analysis-based extraction method using a tool called Sentires from [47], triples containing features, opinions, and sentiments such as $(f, o, s)$ are extracted from all the user-item reviews of $D$. The features extracted form the aspect set $F$ from which explanations are derived from. The opinions are the adjectives which are used to classify the aspect while the sentiment predicted belongs to a binary set of being expressed as either positive or negative i.e. $\{-1, +1\}$. Using these triplets, we create the user-aspect matrix $X \in \mathbb{R}^{m \times r}$. Similarly, we construct item-aspect matrix $Y \in \mathbb{R}^{n \times r}$ where $m = |U|$, $n = |V|$ and $r = |F|$. We adopted the same construction technique as done in [8, 20, 35, 38, 46] as follows:

$$X_{u,f} = \begin{cases} 0 & \text{if u not mentions f} \\ 1 + (N-1)\left(\frac{1-\exp(-t_{u,f})}{1+\exp(-t_{u,f})}\right) & \text{else} \end{cases}$$

$$Y_{v,f} = \begin{cases} 0 & \text{if v not reviewed on f} \\ 1 + \left(\frac{(N-1)}{1+\exp(-t_{v,f} \cdot s_{v,f})}\right) & \text{else} \end{cases}$$

where N is the maximum rating scale from the reviews (typically 5), $t_{u,f}$ is the number of mentions from user $u \in U$ on the feature $f \in F$, $t_{v,f}$ is the number of mentions on item $v \in V$ using feature $f \in F$, and $s_{v,f}$ is the average sentiment polarity of all the $(v, f)$ mentions.

We use these matrices as inputs for learning a black-box recommender $g$ with trainable model weights $\Theta$ which predicts the matching score $s_{u,v} = g(X_u, Y_v \mid \Theta; D)$ for a user $u \in U$ and item $v \in V$ where $X_u$ is the user-feature vector corresponding to $u$ in $X$ and $Y_v$ is the item-feature vector corresponding to $v$ in $Y$. For obtaining the top-$K$ recommendation for a user $u$, we find the first $K$ items which score the highest as per the recommender $g$ as
$$R^u = \underset{v \in V, |R^u|=K}{\arg\max} s_{u,v}.$$

### 3.2 Explanation under Noises

Let the explainability capability of the recommender $g(\Theta; D)$ trained on $D$ be $Q$ under a normal condition. In this paper, we strongly hypothesize that when we perturb the model parameters by $\Delta$ (constrained by a max norm constraint $\epsilon$, see eq. 1), the explainability measure changes to $Q'$. In this study, we plan to characterize the difference $Q - Q'$ under different noise levels $\epsilon$. Our hypothesis is that the difference $Q - Q'$ increases as we increase the noise level $\epsilon$. In addition, we strongly believe that adversarial noises (FGSM) would cause a bigger difference than the random noise counterpart since FGSM noises learn much more against the objective learned by a recommender.

**Table 1: Dataset Statistics**

| Dataset | Users | Items | Features | Reviews | Sparsity(%) |
|---|---|---|---|---|---|
| Electronics | 2,832 | 19,816 | 105 | 53,295 | 0.09497 |
| Yelp | 12,163 | 20,256 | 107 | 510,396 | 0.2072 |

## 4 EXPERIMENTS
### 4.1 Datasets & Preprocessing

We chose two datasets of different scales to conduct our experiments:

- **Amazon Electronics**[1]: E-commerce-based dataset which contains user-provided reviews of electronics purchased on the Amazon platform.
- **Yelp**[2]: Users' reviews are contained for various businesses: restaurants, salons, travel agencies, hotels, etc. across different locations in the world.

In order to improve the density of both datasets, we preprocessed by retaining users with at least 20 reviews for the Yelp dataset and at least 10 reviews for the Amazon dataset. Following previous works [16, 35], we created the testing set as follows: for each user, we keep the last 5 interacted items (positive items) by time and randomly sample 100 items that are not at all interacted by the user (negative items). In table 1, we present the final dataset statistics.

### 4.2 White-box based Perturbation attacks

In order to attack the RS using model-based perturbation methods, we chose two simple attack strategies described below:

- **Random Noise**: Gaussian Noise which is drawn from the Normal distribution $N(0, 1)$. We normalize the added noise using $L_2$ norm[3] and then we are easily able to scale them to a global noise level $\epsilon$.
- **Adversarial Noise**: Let the $Loss(D; \Theta)$ be the loss function of the recommender (predominantly combining the utility and explainability of the recommender). We can optimize for the original model parameters as $\hat{\Theta} = \arg\min_{\Theta} Loss(D; \Theta)$. The adversarial noise $\Delta$ which is added as perturbation into the model is learnt after we learn the original model weights $\hat{\Theta}$. According to eq. 1, since this noise follows a max-norm constraint (where $\epsilon$ is total magnitude of adversarial perturbations and $\|\cdot\|$ is the $L_2$ norm.) and it is intractable to exactly maximize for a recommender loss function in general cases, we follow the optimization technique as done in [19, 36, 43] and optimize the noise inspired by the Fast Gradient Sign Method (FGSM)[18] as:

$$\Delta^* = \arg\max_{\Delta; \|\Delta\| \leq \epsilon} Loss(D; \hat{\Theta} + \Delta) \qquad (1)$$

$$\Delta^* = \epsilon \frac{\Gamma}{\|\Gamma\|} \text{ where } \Gamma = \frac{\partial Loss(D; \hat{\Theta} + \Delta)}{\partial \Delta} \qquad (2)$$

---
[1]https://nijianmo.github.io/amazon/index.html.
[2]https://www.yelp.com/dataset.
[3]Embeddings are normalized and the max-norm constraint is enforced per each column.

### 4.3 Models & Training Setting

For verifying our hypothesis that existent explainable models are vulnerable to external attacks, we pick SOTA feature-based explainable recommenders chosen based on the presence and involvement of explicit and/or hidden factors in both the recommendation and explanation procedures. The models are described as follows:

- **CER: Counterfactual Explainable Recommendation** [35]: The top $K$ recommendations are learned from a black box neural network model comprised of two hidden layers with $X$ and $Y$ as inputs. Then, the explanations for the top $K$ items for any user are the most minimal counterfactual changes done to the item features space such that these changes are responsible for the item not being recommended in the top $K$ list to this user. We chose to perturb all the hidden layers of the recommender neural network model since these are responsible for learning from the features for predicting recommendations. We also provide personalized explanations per user which then only explains from the features mentioned by the user in all their reviews.

- **A2CF: Aspect Aware Collaborative Filtering** [8]: This paper predicts the missing user-feature and item-feature values within $X$ and $Y$ using a residual neural network by learning user, item, and feature embeddings. We remove the item-item similarity learning so that the model is capable to explain all the user-item pairs. Since we mainly wanted to focus on studying the impact of direct explicit factors responsible for both recommendation and explanation in this model, we chose to perturb all the three embeddings: user, item, and aspect. In addition, we also perturbed the projection weight used for predicting matching scores (via Bayesian Pairwise Ranking) for user-item pairs.

- **EFM: Explicit Factor Modeling** [46]: This work is heavily based on matrix factorization techniques by decomposing three matrices: user-item interaction matrix, user-feature matrix $X$, and item-feature matrix $Y$ into smaller rank matrices learning with both explicit and hidden factors. For ensuring an even-handed influence of both explicit and hidden factors, we set the hidden dimension as the same for both. For this model, we perturbed only the explicit factor matrices $U_1, U_2, V$ since they are the only factors that are responsible for both recommendation and explanation. Additionally, since this method has a closed-form solution (unlike the other models considered which have employed gradient descent for optimization) for finding the optimal parameters, we only perturbed using the random noise method.

*Training setting:* For all the cases, the models are trained until convergence with a batch size of 32. We chose the best hyperparameters for each model by grid search. We set the learning rate as 0.001 and Stochastic Gradient Descent was used for optimizing all the gradient-descent based models. The FGSM-based attack models were trained in the same conditions as the vanilla model was trained. We first provided the top-$K$ = 5 recommendation lists for each user and then we explain using the top-$E$ = 5 features. We also chose $\epsilon$ from [0,1] for Yelp and [0,2] for Electronics datasets.

### 4.4 Evaluation

For evaluating the recommendation quality, we ended up choosing the most common metric of Normalized Discounted Cumulative Gain (NDCG). However, in order to gauge feature-level explanations, we utilized Feature-level Precision, Recall, and F1 (harmonic mean of Precision and Recall) scores of the explanations by comparing them with the golden truth features found in the reviews of the user-item interaction, which have been mentioned with positive sentiment as suggested by papers [34, 35, 42, 48]. We evaluate all the explained samples for which a review actually exists (positive interactions by a user) and report the average metric scores across all such samples.

## 5 RESULTS

From the results, there is a clear inference that the explanation performance drops heavily on increasing noises for almost all the models which verifies our hypothesis that feature-aware explainable recommenders are vulnerable and hence explanation methods within RS are prone to attacks. We can also clearly conclude that the potency of adversarial attacks (see figure 2) is far superior when compared to random attacks. This is because the FGSM based adversarial attacks learn model perturbations by optimizing against the original objective of the recommender (by learning in the opposite direction of the gradients) which causes the model to inflict stronger behavior change and expose more vulnerabilities within the model.

### 5.1 Lack of Generalization in the Explanations

Based on the observed results, we can infer that the explanations are not generalized and robust across changing scenarios across all the recommenders. This is because the recommenders do not generalize well in capturing the exact features that correspond with the user preferences for a particular recommendation. As the noises are added into the model, the predicted outcome of the model becomes much more incorrect compared to the original vanilla model case, leading to a wrong identification of the most contributing features to explain the outcome. This implies that the explanation provided by all the models in general is not reliable indicating a vulnerability within the models. The recommendation quality drops as the noise increases leading to the incorrect features used for explaining the recommendation provided to the user and hence leading to an eventual drop in the global explainability of the system. We can observe this vulnerability in both the datasets (see tables 2 and 3) implying that the attacked models lose their capability to provide stable explanations for any suggested item on average and this trend (in figure 2) just deteriorates for increasing $\epsilon$.

### 5.2 Model architecture: Explicit vs hidden factors?

The main reasons we suspect for observing this vulnerability are the impact of hidden factors and the explicit factors involved in the recommendation and explanation tasks of the recommenders. We deduce that the impact of the hidden and explicit factors depends on

Table 2: Vulnerability of Explainable Recommenders on Electronics

| | Noise Level ($\epsilon$) | Random | | | | FGSM | | | |
|---|---|---|---|---|---|---|---|---|---|
| | | NDCG@100 | Pr@5,5 | Re@5,5 | F1@5,5 | NDCG@100 | Pr@5,5 | Re@5,5 | F1@5,5 |
| **CER** | 0 (vanilla) | 0.4889 | 0.04115 | 0.1331 | 0.05928 | 0.4889 | 0.04115 | 0.1331 | 0.05928 |
| | 0.5 | 0.4891 | 0.04021 | 0.1309 | 0.05805 | 0.4666 | 0.01733 | 0.05386 | 0.02481 |
| | 1 | 0.4886 | 0.04016 | 0.1304 | 0.05796 | 0.4312 | 0.02046 | 0.05964 | 0.02866 |
| | 1.5 | 0.4882 | 0.04123 | 0.1351 | 0.05955 | 0.4081 | 0.01658 | 0.05635 | 0.02445 |
| | 2 | 0.4878 | 0.04164 | 0.1363 | 0.06009 | 0.3993 | 0.02069 | 0.08333 | 0.03223 |
| | Noise Level ($\epsilon$) | Random | | | | FGSM | | | |
| | | NDCG@100 | Pr@5,5 | Re@5,5 | F1@5,5 | NDCG@100 | Pr@5,5 | Re@5,5 | F1@5,5 |
| **A2CF** | 0 (vanilla) | 0.4059 | 0.02996 | 0.1026 | 0.04412 | 0.4059 | 0.02996 | 0.1026 | 0.04412 |
| | 0.5 | 0.4047 | 0.02537 | 0.083 | 0.03685 | 0.3991 | 0.02121 | 0.07244 | 0.03105 |
| | 1 | 0.3973 | 0.02193 | 0.07458 | 0.03225 | 0.3927 | 0.01813 | 0.06205 | 0.0265 |
| | 1.5 | 0.3881 | 0.02047 | 0.06941 | 0.02993 | 0.3728 | 0.00879 | 0.02513 | 0.01229 |
| | 2 | 0.3819 | 0.01774 | 0.05913 | 0.02601 | 0.3603 | 0.00619 | 0.02002 | 0.00874 |
| | Noise Level ($\epsilon$) | Random | | | | | | | |
| | | NDCG@100 | Pr@5,5 | | Re@5,5 | | F1@5,5 | | |
| **EFM** | 0 (vanilla) | 0.4319 | 0.05988 | | 0.219 | | 0.08996 | | |
| | 0.5 | 0.4015 | 0.05159 | | 0.1797 | | 0.07634 | | |
| | 1 | 0.3961 | 0.02986 | | 0.1087 | | 0.04476 | | |
| | 1.5 | 0.3864 | 0.02273 | | 0.08107 | | 0.0338 | | |
| | 2 | 0.38 | 0.01913 | | 0.06937 | | 0.02839 | | |

Table 3: Vulnerability of Explainable Recommenders on Yelp

| | Noise Level ($\epsilon$) | Random | | | | FGSM | | | |
|---|---|---|---|---|---|---|---|---|---|
| | | NDCG@100 | Pr@5,5 | Re@5,5 | F1@5,5 | NDCG@100 | Pr@5,5 | Re@5,5 | F1@5,5 |
| **CER** | 0 (vanilla) | 0.5365 | 0.0139 | 0.0695 | 0.02317 | 0.5365 | 0.0139 | 0.0695 | 0.02317 |
| | 0.2 | 0.5362 | 0.01418 | 0.07088 | 0.02363 | 0.5155 | 0.00985 | 0.04927 | 0.01642 |
| | 0.4 | 0.5361 | 0.01429 | 0.07146 | 0.02382 | 0.508 | 0.00891 | 0.0446 | 0.01485 |
| | 0.6 | 0.5358 | 0.01409 | 0.07046 | 0.02349 | 0.4846 | 0.00971 | 0.04854 | 0.01618 |
| | 0.8 | 0.5354 | 0.01411 | 0.07054 | 0.02351 | 0.4347 | 0.01001 | 0.05002 | 0.01667 |
| | 1 | 0.535 | 0.0135 | 0.06748 | 0.02249 | 0.3441 | 0.00952 | 0.04762 | 0.01587 |
| | Noise Level ($\epsilon$) | Random | | | | FGSM | | | |
| | | NDCG@100 | Pr@5,5 | Re@5,5 | F1@5,5 | NDCG@100 | Pr@5,5 | Re@5,5 | F1@5,5 |
| **A2CF** | 0 (vanilla) | 0.6454 | 0.01225 | 0.06126 | 0.02042 | 0.6454 | 0.01225 | 0.06126 | 0.02042 |
| | 0.2 | 0.6434 | 0.01256 | 0.0628 | 0.02093 | 0.6456 | 0.01112 | 0.0556 | 0.01853 |
| | 0.4 | 0.6378 | 0.01258 | 0.06289 | 0.02096 | 0.6419 | 0.00947 | 0.04734 | 0.01578 |
| | 0.6 | 0.6289 | 0.01229 | 0.06144 | 0.02048 | 0.6438 | 0.00916 | 0.04581 | 0.01527 |
| | 0.8 | 0.6181 | 0.01235 | 0.06177 | 0.02059 | 0.6462 | 0.00933 | 0.04664 | 0.01555 |
| | 1 | 0.6043 | 0.01215 | 0.06073 | 0.02024 | 0.6361 | 0.00901 | 0.04503 | 0.01501 |
| | Noise Level ($\epsilon$) | Random | | | | | | | |
| | | NDCG@100 | Pr@5,5 | | Re@5,5 | | F1@5,5 | | |
| **EFM** | 0 (vanilla) | 0.3127 | 0.01849 | | 0.09244 | | 0.03081 | | |
| | 0.2 | 0.231 | 0.01805 | | 0.09024 | | 0.03008 | | |
| | 0.4 | 0.14 | 0.018 | | 0.09 | | 0.03 | | |
| | 0.6 | 0.2809 | 0.01659 | | 0.08297 | | 0.02766 | | |
| | 0.8 | 0.2844 | 0.01413 | | 0.07064 | | 0.02355 | | |
| | 1 | 0.2878 | 0.0142 | | 0.07101 | | 0.02367 | | |

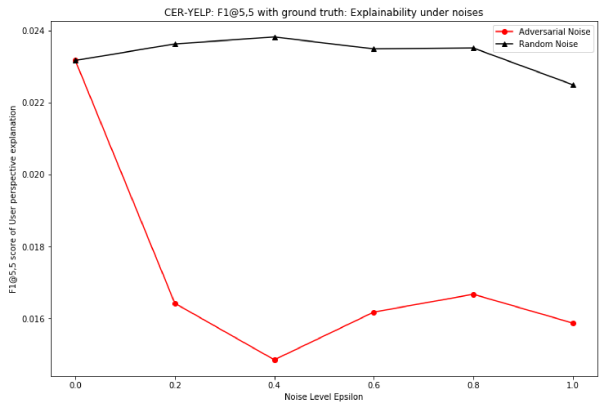
(a) CER F1@5,5 on Yelp

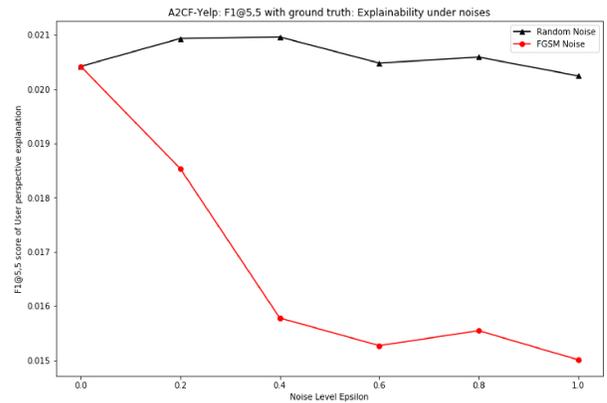
(b) A2CF F1@5,5 on Yelp

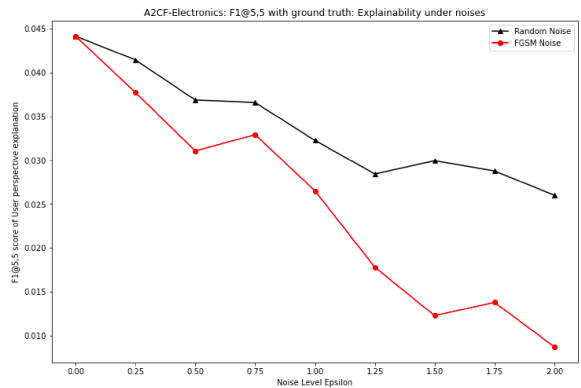
(c) A2CF F1@5,5 on Electronics

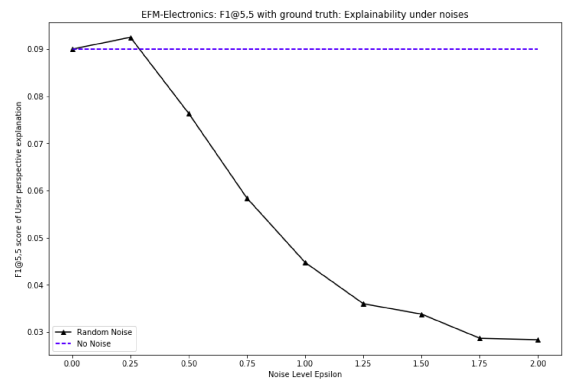
(d) EFM F1@5,5 on Electronics

**Figure 2: Vulnerability of feature-aware explainable recommenders under different noises. The black plots correspond to the random noise while the red plots correspond to the adversarial noise. The blue line in EFM represents the no-noise (for vanilla) value as a dotted line. The X-axis is the noise level $\epsilon$ and the Y-axis is F1@5,5**

the scale of dataset used for training. In the following sub sections, we would discuss the reasons for the noticed effect on both the datasets used in the experiments.

5.2.1 **Effect on Electronics**. On Electronics, we discover that the explicit factor-based models are much more vulnerable leading the model to lose the explainability capability as we increase $\epsilon$. A2CF learns three sets of embeddings for both tasks and although it uses a residual neural network for reconstruction tasks, the number of explicit factors learnt are much greater than the number of learnt hidden parameters. Therefore, this describes why A2CF is the weakest when trained on the Electronics dataset. We notice a similar impact in the case of the EFM model, which is primarily a pure matrix factorization-based method, where just adding plain Gaussian noises yields a huge drop in explainability. This also accounts for the fact that CER, being a neural network model, does not experience a severe drop in explainability for a small dataset. So, we conclude that a smaller dataset is heavily influenced by the explicit factors within both recommendation and explanation tasks.

5.2.2 **Effect on Yelp**. However, the reasons for the observed vulnerabilities are totally different when we inspect a larger dataset. When we conducted the experiments on Yelp, we observe the largest decrease in explainability occurs in the CER model. We suspect that the root cause for this vulnerability is due to the complete neural network architecture adopted by CER. Typically, neural network models are vulnerable to noises [7, 25, 27] for large-scale datasets and this could possibly be the reason for the CER model to experience the highest drop in explainability. However, A2CF and EFM models, both being more explicit factor-based, experience lesser impact on explainability compared to a traditional neural network model when perturbed by increasing levels of noise. So, an increasing scale of the dataset yields to a much more pronounced effect of decrease in explanations in neural network-based models (by virtue of possessing more hidden factors) than factorization-based (which possess more explicit factors) ones.

## 6 CONCLUSIONS & FUTURE WORK

In this study, we explored a fresh problem of vulnerability of existent feature-aware explainable recommenders and we conducted extensive experiments of different feature-aware explainable RS under varying noise levels $\epsilon$ by perturbing the models with two different kinds of noises: random and adversarial. From our experiments, we conclude all the models are vulnerable to external attacks, implying worse identification of reasons for the predicted outcome thus decreasing explainability. We also provided attributed the noticed behavior with the presence of explicit and hidden factors. We also discuss how these factors play a huge role as the size of the dataset increases.

While we present this fresh empirical study regarding various explainable methods in RS, we also note that there is much scope for future work and developments. The most important extension to this work would be developing newer explainer methods in RS that are capable of generating explanations that are robust in general and can be relied upon by consumers. We also highlight that different types of stability such as generalized explainability across different domains could be analyzed, besides security-based stability as in this study. Robust explanations are more than a necessity in RS since it possesses serious consequences for both consumers and developers. Finally, we strongly emphasize that the field of robust explanations deserve a much broader study within the field of explainable AI and particularly in RS.

## REFERENCES

[1] Julius Adebayo, Justin Gilmer, Michael Muelly, Ian Goodfellow, Moritz Hardt, and Been Kim. 2018. Sanity checks for saliency maps. *Advances in neural information processing systems* 31 (2018).

[2] Chirag Agarwal, Nari Johnson, Martin Pawelczyk, Satyapriya Krishna, Eshika Saxena, Marinka Zitnik, and Himabindu Lakkaraju. 2022. Rethinking Stability for Attribution-based Explanations. http://arxiv.org/abs/2203.06877 arXiv:2203.06877 [cs].

[3] Chirag Agarwal, Marinka Zitnik, and Himabindu Lakkaraju. 2022. Probing GNN Explainers: A Rigorous Theoretical and Empirical Analysis of GNN Explanation Methods. http://arxiv.org/abs/2106.09078 arXiv:2106.09078 [cs].

[4] David Alvarez-Melis and Tommi S. Jaakkola. 2018. On the Robustness of Interpretability Methods. http://arxiv.org/abs/1806.08049 arXiv:1806.08049 [cs, stat].

[5] David Alvarez-Melis and Tommi S. Jaakkola. 2018. Towards Robust Interpretability with Self-Explaining Neural Networks. http://arxiv.org/abs/1806.07538 arXiv:1806.07538 [cs, stat].

[6] Umang Bhatt, Alice Xiang, Shubham Sharma, Adrian Weller, Ankur Taly, Yunhan Jia, Joydeep Ghosh, Ruchir Puri, José M. F. Moura, and Peter Eckersley. 2020. Explainable machine learning in deployment. In *Proceedings of the 2020 Conference on Fairness, Accountability, and Transparency*. ACM, Barcelona Spain, 648–657. https://doi.org/10.1145/3351095.3375624

[7] Ali Borji and Sikun Lin. 2019. White Noise Analysis of Neural Networks. arXiv:1912.12106 [cs.CV]

[8] Tong Chen, Hongzhi Yin, Guanhua Ye, Zi Huang, Yang Wang, and Meng Wang. 2020. Try This Instead: Personalized and Interpretable Substitute Recommendation. In *Proceedings of the 43rd International ACM SIGIR Conference on Research and Development in Information Retrieval*. ACM, Virtual Event China, 891–900. https://doi.org/10.1145/3397271.3401042

[9] Zhiyong Cheng, Ying Ding, Xiangnan He, Lei Zhu, Xuemeng Song, and Mohan Kankanhalli. 2018. A^3NCF: An Adaptive Aspect Attention Model for Rating Prediction. In *Proceedings of the Twenty-Seventh International Joint Conference on Artificial Intelligence*. International Joint Conferences on Artificial Intelligence Organization, Stockholm, Sweden, 3748–3754. https://doi.org/10.24963/ijcai.2018/521

[10] Ronky Francis Doh, Conghua Zhou, John Kingsley Arthur, Isaac Tawiah, and Benjamin Doh. 2022. A Systematic Review of Deep Knowledge Graph-Based Recommender Systems, with Focus on Explainable Embeddings. *Data* 7, 7 (July 2022), 94. https://doi.org/10.3390/data7070094

[11] Ann-Kathrin Dombrowski, Maximilian Alber, Christopher J. Anders, Marcel Ackermann, Klaus-Robert Müller, and Pan Kessel. 2019. Explanations can be manipulated and geometry is to blame. http://arxiv.org/abs/1906.07983 arXiv:1906.07983 [cs, stat].

[12] Christian Etmann, Sebastian Lunz, Peter Maass, and Carola-Bibiane Schönlieb. 2019. On the Connection Between Adversarial Robustness and Saliency Map Interpretability. http://arxiv.org/abs/1905.04172 arXiv:1905.04172 [cs, stat].

[13] Wenqi Fan, Wei Jin, Xiaorui Liu, Han Xu, Xianfeng Tang, Suhang Wang, Qing Li, Jiliang Tang, Jianping Wang, and Charu Aggarwal. 2022. Jointly Attacking Graph Neural Network and its Explanations. http://arxiv.org/abs/2108.03388 arXiv:2108.03388 [cs].

[14] Wenqi Fan, Xiangyu Zhao, Xiao Chen, Jingran Su, Jingtong Gao, Lin Wang, Qidong Liu, Yiqi Wang, Han Xu, Lei Chen, et al. 2022. A Comprehensive Survey on Trustworthy Recommender Systems. *arXiv preprint arXiv:2209.10117* (2022).

[15] Yingqiang Ge, Shuchang Liu, Zuohui Fu, Juntao Tan, Zelong Li, Shuyuan Xu, Yunqi Li, Yikun Xian, and Yongfeng Zhang. 2022. A survey on trustworthy recommender systems. *arXiv preprint arXiv:2207.12515* (2022).

[16] Yingqiang Ge, Juntao Tan, Yan Zhu, Yinglong Xia, Jiebo Luo, Shuchang Liu, Zuohui Fu, Shijie Geng, Zelong Li, and Yongfeng Zhang. 2022. Explainable Fairness in Recommendation. In *Proceedings of the 45th International ACM SIGIR Conference on Research and Development in Information Retrieval*. ACM, Madrid Spain, 681–691. https://doi.org/10.1145/3477495.3531973

[17] Amirata Ghorbani, Abubakar Abid, and James Zou. 2019. Interpretation of Neural Networks Is Fragile. *Proceedings of the AAAI Conference on Artificial Intelligence* 33, 01 (July 2019), 3681–3688. https://doi.org/10.1609/aaai.v33i01.33013681

[18] Ian J. Goodfellow, Jonathon Shlens, and Christian Szegedy. 2014. Explaining and Harnessing Adversarial Examples. (2014). https://doi.org/10.48550/ARXIV.1412.6572 Publisher: arXiv Version Number: 3.

[19] Xiangnan He, Zhankui He, Xiaoyu Du, and Tat-Seng Chua. 2018. Adversarial Personalized Ranking for Recommendation. In *The 41st International ACM SIGIR Conference on Research & Development in Information Retrieval*. 355–364. https://doi.org/10.1145/3209978.3209981 arXiv:1808.03908 [cs, stat].

[20] Yunfeng Hou, Ning Yang, Yi Wu, and Philip S. Yu. 2019. Explainable recommendation with fusion of aspect information. *World Wide Web* 22, 1 (Jan. 2019), 221–240. https://doi.org/10.1007/s11280-018-0558-1

[21] Chao Huang, Huance Xu, Yong Xu, Peng Dai, Lianghao Xia, Mengyin Lu, Liefeng Bo, Hao Xing, Xiaoping Lai, and Yanfang Ye. 2021. Knowledge-aware Coupled Graph Neural Network for Social Recommendation. *Proceedings of the AAAI Conference on Artificial Intelligence* 35, 5 (May 2021), 4115–4122. https://doi.org/10.1609/aaai.v35i5.16533

[22] Beomsu Kim, Junghoon Seo, and Taegyun Jeon. 2019. Bridging Adversarial Robustness and Gradient Interpretability. http://arxiv.org/abs/1903.11626 arXiv:1903.11626 [cs, stat].

[23] Trung-Hoang Le and Hady W. Lauw. 2021. Explainable Recommendation with Comparative Constraints on Product Aspects. In *Proceedings of the 14th ACM International Conference on Web Search and Data Mining*. ACM, Virtual Event Israel, 967–975. https://doi.org/10.1145/3437963.3441754

[24] Lei Li, Yongfeng Zhang, and Li Chen. 2021. Personalized Transformer for Explainable Recommendation. http://arxiv.org/abs/2105.11601 arXiv:2105.11601 [cs].

[25] Mengchen Liu, Shixia Liu, Hang Su, Kelei Cao, and Jun Zhu. 2018. Analyzing the Noise Robustness of Deep Neural Networks. arXiv:1810.03913 [cs.LG]

[26] Yanzhang Lyu, Hongzhi Yin, Jun Liu, Mengyue Liu, Huan Liu, and Shizhuo Deng. 2021. Reliable Recommendation with Review-level Explanations. In *2021 IEEE 37th International Conference on Data Engineering (ICDE)*. IEEE, Chania, Greece, 1548–1558. https://doi.org/10.1109/ICDE51399.2021.00137

[27] Marko Mihajlović and Nikola Popović. 2018. Fooling a neural network with common adversarial noise. In *2018 19th IEEE Mediterranean Electrotechnical Conference (MELECON)*. 293–296. https://doi.org/10.1109/MELCON.2018.8379110

[28] Ian E. Nielsen, Dimah Dera, Ghulam Rasool, Nidhal Bouaynaya, and Ravi P. Ramachandran. 2022. Robust Explainability: A Tutorial on Gradient-Based Attribution Methods for Deep Neural Networks. *IEEE Signal Processing Magazine* 39, 4 (July 2022), 73–84. https://doi.org/10.1109/MSP.2022.3142719 arXiv:2107.11400 [cs].

[29] Sicheng Pan, Dongsheng Li, Hansu Gu, Tun Lu, Xufang Luo, and Ning Gu. 2022. Accurate and Explainable Recommendation via Review Rationalization. In *Proceedings of the ACM Web Conference 2022*. ACM, Virtual Event, Lyon France, 3092–3101. https://doi.org/10.1145/3485447.3512029

[30] Xiao Sha, Zhu Sun, and Jie Zhang. 2021. Hierarchical Attentive Knowledge Graph Embedding for Personalized Recommendation. *Electronic Commerce Research and Applications* 48 (July 2021), 101071. https://doi.org/10.1016/j.elerap.2021.101071 arXiv:1910.08288 [cs].

[31] Avanti Shrikumar, Peyton Greenside, and Anshul Kundaje. 2019. Learning Important Features Through Propagating Activation Differences. arXiv:1704.02685 [cs.CV]

[32] Dylan Slack, Sophie Hilgard, Emily Jia, Sameer Singh, and Himabindu Lakkaraju. 2020. Fooling LIME and SHAP: Adversarial Attacks on Post hoc Explanation Methods. http://arxiv.org/abs/1911.02508 arXiv:1911.02508 [cs, stat].

[33] Akshayvarun Subramanya, Vipin Pillai, and Hamed Pirsiavash. 2019. Fooling Network Interpretation in Image Classification. http://arxiv.org/abs/1812.02843 arXiv:1812.02843 [cs].

[34] Chang-You Tai, Liang-Ying Huang, Chien-Kun Huang, and Lun-Wei Ku. 2021. User-centric path reasoning towards explainable recommendation. In *Proceedings of the 44th International ACM SIGIR Conference on Research and Development in Information Retrieval*. 879–889.

[35] Juntao Tan, Shuyuan Xu, Yingqiang Ge, Yunqi Li, Xu Chen, and Yongfeng Zhang. 2021. Counterfactual Explainable Recommendation. In *Proceedings of the 30th ACM International Conference on Information & Knowledge Management*. ACM, Virtual Event Queensland Australia, 1784–1793. https://doi.org/10.1145/3459637.3482420

[36] Thanh Tran, Renee Sweeney, and Kyumin Lee. 2019. Adversarial Mahalanobis Distance-based Attentive Song Recommender for Automatic Playlist Continuation. In *Proceedings of the 42nd International ACM SIGIR Conference on Research and Development in Information Retrieval*. ACM, Paris France, 245–254. https://doi.org/10.1145/3331184.3331234

[37] Alexandra Vultureanu-Albişi and Costin Bădică. 2022. A survey on effects of adding explanations to recommender systems. *Concurrency and Computation: Practice and Experience* 34, 20 (Sept. 2022). https://doi.org/10.1002/cpe.6834

[38] Nan Wang, Hongning Wang, Yiling Jia, and Yue Yin. 2018. Explainable Recommendation via Multi-Task Learning in Opinionated Text Data. In *The 41st International ACM SIGIR Conference on Research & Development in Information Retrieval*. 165–174. https://doi.org/10.1145/3209978.3210010 arXiv:1806.03568 [cs].

[39] Shoujin Wang, Xiuzhen Zhang, Yan Wang, Huan Liu, and Francesco Ricci. 2022. Trustworthy Recommender Systems. http://arxiv.org/abs/2208.06265 arXiv:2208.06265 [cs].

[40] Bingbing Wen, Yunhe Feng, Yongfeng Zhang, and Chirag Shah. 2022. Towards Generating Robust, Fair, and Emotion-Aware Explanations for Recommender Systems. http://arxiv.org/abs/2208.08017 arXiv:2208.08017 [cs].

[41] Yikun Xian, Tong Zhao, Jin Li, Jim Chan, Andrey Kan, Jun Ma, Xin Luna Dong, Christos Faloutsos, George Karypis, S. Muthukrishnan, and Yongfeng Zhang. 2021. EX3: Explainable Attribute-aware Item-set Recommendations. In *Fifteenth ACM Conference on Recommender Systems*. ACM, Amsterdam Netherlands, 484–494. https://doi.org/10.1145/3460231.3474240

[42] Ning Yang, Yuchi Ma, Li Chen, and Philip S Yu. 2020. A meta-feature based unified framework for both cold-start and warm-start explainable recommendations. *World Wide Web* 23 (2020), 241–265.

[43] Feng Yuan, Lina Yao, and Boualem Benatallah. 2019. Adversarial Collaborative Neural Network for Robust Recommendation. In *Proceedings of the 42nd International ACM SIGIR Conference on Research and Development in Information Retrieval*. ACM, Paris France, 1065–1068. https://doi.org/10.1145/3331184.3331321

[44] Xinyang Zhang, Ningfei Wang, Hua Shen, Shouling Ji, Xiapu Luo, and Ting Wang. 2020. Interpretable deep learning under fire. In *29th {USENIX} security symposium ({USENIX} security 20)*.

[45] Yongfeng Zhang and Xu Chen. 2020. Explainable Recommendation: A Survey and New Perspectives. *Foundations and Trends® in Information Retrieval* 14, 1 (2020), 1–101. https://doi.org/10.1561/1500000066

[46] Yongfeng Zhang, Guokun Lai, Min Zhang, Yi Zhang, Yiqun Liu, and Shaoping Ma. 2014. Explicit factor models for explainable recommendation based on phrase-level sentiment analysis. In *Proceedings of the 37th international ACM SIGIR conference on Research & development in information retrieval*. ACM, Gold Coast Queensland Australia, 83–92. https://doi.org/10.1145/2600428.2609579

[47] Yongfeng Zhang, Haochen Zhang, Min Zhang, Yiqun Liu, and Shaoping Ma. 2014. Do users rate or review?: boost phrase-level sentiment labeling with review-level sentiment classification. In *Proceedings of the 37th international ACM SIGIR conference on Research & development in information retrieval*. ACM, Gold Coast Queensland Australia, 1027–1030. https://doi.org/10.1145/2600428.2609501

[48] Yao Zhou, Haonan Wang, Jingrui He, and Haixun Wang. 2021. From Intrinsic to Counterfactual: On the Explainability of Contextualized Recommender Systems. http://arxiv.org/abs/2110.14844 arXiv:2110.14844 [cs].